\documentclass[twocolumn,english,aps,prl,showpacs,amssymb,amsfonts,superscriptaddress]{revtex4-1}
\usepackage[T1]{fontenc}
\usepackage[latin9]{inputenc}
\usepackage{amsmath}
\usepackage{graphicx}
\usepackage{amssymb}
\usepackage{color}
\usepackage{enumerate}
\usepackage{times}

\usepackage[colorlinks=true,citecolor=blue,linkcolor=blue]{hyperref}

\newcommand{\beq}{\begin{equation}}
\newcommand{\eeq}{\end{equation}}
\newcommand{\bea}{\begin{eqnarray}}
\newcommand{\eea}{\end{eqnarray}}

\definecolor{purple}{rgb}{0.5,0,0.5}

\def\ket#1{{\left|#1\right\rangle}}

\def\bk{\boldsymbol{k}}

\def\ba{\boldsymbol{a}}

\def\bK{\boldsymbol{K}}

\def\be{\begin{eqnarray}}
\def\ee{\end{eqnarray}}

\newcommand{\bs}{\boldsymbol}

\def \a{{\alpha}}

\def \D{{\Delta}}

\def \ket#1{{\,|\,#1\,\rangle\,}}

\begin{document}

\title{Chiral Bosonic Mott Insulator on the Frustrated Triangular Lattice}
\author {Michael P. Zaletel}
\affiliation{Department of Physics, University of California, Berkeley, CA 94720, USA}
\author {S. A. Parameswaran}
\affiliation{Department of Physics, University of California, Berkeley, CA 94720, USA}
\affiliation{Department of Physics and Astronomy, University of California, Irvine, CA 92697, USA}
\author {Andreas R\"uegg}
\affiliation{Department of Physics, University of California, Berkeley, CA 94720, USA}
\affiliation{Theoretische Physik, Wolfgang-Pauli-Strasse 27, ETH Z\"urich, CH-8093 Z\"urich, Switzerland}
\author {Ehud Altman}
\affiliation{Department of Physics, University of California, Berkeley, CA 94720, USA}
\affiliation{Department of Condensed Matter Physics, Weizmann Institute of Science, Rehovot 76100, Israel}

\date{\today}

\begin{abstract}
We study the superfluid and insulating phases of interacting bosons on the triangular lattice with an inverted dispersion, corresponding to frustrated hopping between sites. The resulting single-particle dispersion has multiple minima at nonzero wavevectors in momentum space, in contrast to the unique zero-wavevector minimum of the unfrustrated problem. As a consequence, the superfluid phase is unstable against developing additonal chiral order that breaks time reversal ($\mathcal{T}$) and parity ($\mathcal{P}$)  symmetries by forming a condensate at nonzero wavevector.  We demonstrate that the loss of superfluidity can lead to an even  more exotic phase, the chiral Mott insulator, with nontrivial current order that breaks $\mathcal{T},\mathcal{P}$.
 These results are obtained via variational estimates, as well as a combination of bosonization and DMRG of triangular ladders, which taken together permit a fairly complete characterization of the phase diagram. We discuss the relevance of these phases to optical lattice experiments, as well as signatures of chiral symmetry breaking in time-of-flight images. 
\end{abstract}
\maketitle

The Mott insulating phase of ultracold atoms in an optical lattice is the simplest 
example of a ground state of uncondensed bosons and, to date, the only one demonstrated experimentally~\cite{Greiner2002}. This phase however is not {\it intrinsically} quantum-mechanical as it can be adiabatically connected to a ``classical'' state of decoupled sites with definite occupation. Therefore, it is natural to ask how bosons can insulate while retaining non-trivial quantum correlations
\cite{*[{Crystal symmetry can enforce quantum structure; see for instance  }] [{ and }] KagomeWannier, *HoneycombVoronoi} under realistic conditions.
A compelling setting where this question may be addressed is in
optical lattice systems that realize complex superfluid states~\cite{Wirth2010,Olschlager2011,Struck2011}. In these experiments, 
 a special band dispersion -- achieved via meta-stable occupation of higher bands~\cite{Wirth2010,Olschlager2011} or by rapid lattice modulation~\cite{Struck2011} --  leads to condensation at finite non-trivial momenta. The resulting superfluids 
 spontaneously break time reversal and crystalline space group symmetries. The possibility of  nontrivial insulating behavior then turns on the manner in which superfluidity is lost 
 as the lattice depth is increased at fixed, integer filling.

 In one route, all symmetries are restored simultaneously and the system transitions directly to a featureless Mott insulator.
 A more natural and intriguing possibility, however, is that time reversal symmetry remains broken across the superfluid transition as $U(1)$ phase symmetry is restored. This scenario implies the existence of a correlated time reversal-breaking insulator: a {\it chiral} Mott insulator. Time reversal symmetry is restored only for a deeper lattice, via a second transition.

A similar scenario 
was proposed some time ago for the classical, temperature-tuned transition from a two dimensional time-reversal breaking superfluid to a thermal gas~\cite{Lee1984}. Monte Carlo simulations show that while  superfluidity is lost at the usual Berezinskii-Kosterlitz-Thouless (BKT) transition, time reversal symmetry is restored only at a higher temperature Ising transition. The intermediate chiral liquid phase is the classical analogue of a chiral Mott insulator.
 Since the two dimensional classical problem is formally equivalent to a one dimensional quantum system at zero temperature it is natural to expect that a chiral Mott ground state can be stabilized, at least in one dimension. Indeed, Dhar \emph{et. al.}~\cite{Dhar2012,Dhar2013} examined just such a state on a two-leg square ladder. A closely related gapped spin-current state was recently considered as a candidate for the 1/3 magnetization plateau of a highly anisotropic triangular antiferromagnet~\cite{Chubukov:2013}. In spite of this recent activity, whether a chiral Mott phase can exist in an isotropic system in two or higher dimensions remains open.

\begin{figure}[t]
\includegraphics[width=0.8\columnwidth]{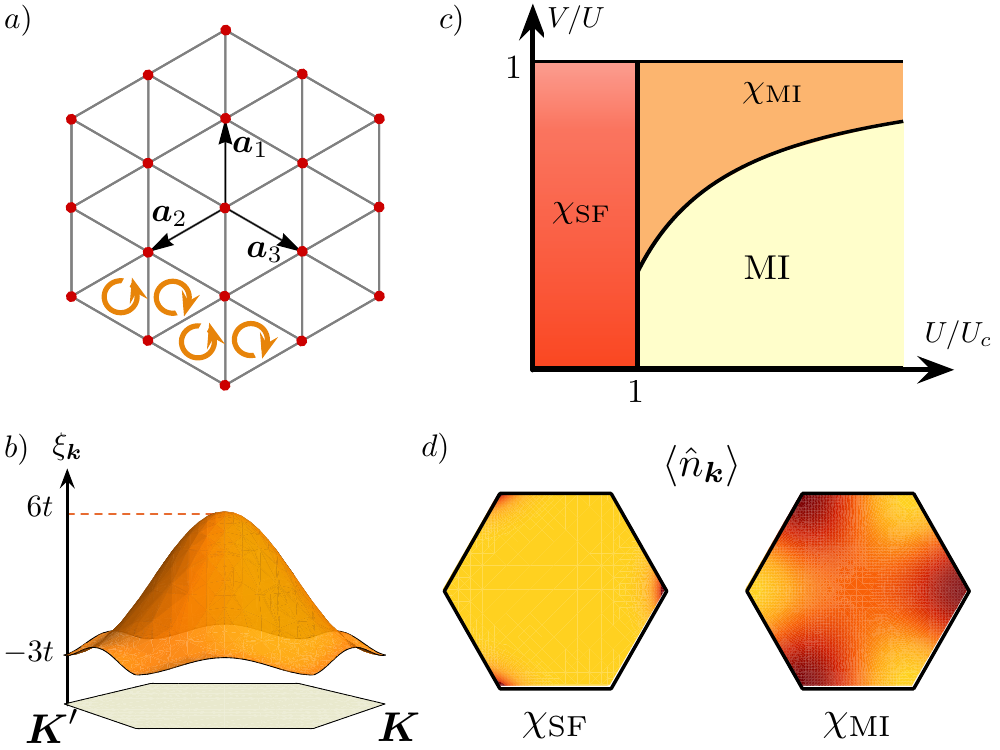}
\caption{{\bf \label{fig:varphasediag} Bosons on the Frustrated Triangular Lattice.} (a) Lattice, coordinate system and sample current pattern in the $\chi$MI; (b)  single-particle dispersion $\xi_{\bk}$, with minima at the $\bK,\bK'$ points of the BZ; (c) Variational mean-field phase diagram showing $\chi$SF, $\chi$MI and MI phases tuned by the on site repulsion $U$ and nearest neighbor repulsion $V$; (d) Momentum distribution $\langle\hat{n}_{\bk}\rangle$ for the chiral phases.}
\end{figure}

Here, we answer this question in the affirmative, by investigating 
an interacting boson model with a {\em positive} nearest-neighbor hopping amplitude $t$ on the triangular lattice. Because of the positive hopping, the band minima are at the inequivalent  momenta $\bK$ and ${\bK}'=-{\bK}$ at the corners of the hexagonal Brillouin zone (BZ). A condensate established at either of these momenta breaks symmetry under time reversal, parity, and $60^{\textdegree}$ rotations,  while preserving the discrete lattice translational symmetry: it is a chiral superfluid.     
We map out the phase diagram of this model in Fig. \ref{fig:varphasediag}  as a function of interaction parameters using a family of variational wave functions that can capture the chiral Mott and superfluid phases, as well as  the trivial Mott insulator. We discuss the signature of the three phases in time-of-flight measurements. 

To supplement the variational analysis we also consider a reduction of the problem to frustrated triangular ladders. First, we recover all three phases on a four-leg ladder within an infinite density-matrix renormalization group (iDMRG) analysis~\cite{White-1992, McCulloch-2008, Kjall2013}. We then study a two-leg ladder which permits  a long-wavelength analysis of the phase diagram via bosonization, and where iDMRG simulations can be performed to sufficient precision to extract critical behavior at the transitions. 
This significantly bolsters the results of the $d=2$ mean-field theory. We close by discussing the relevance of our results to various systems.
 
{\it Variational Phase Diagram.---} We begin with an extended Bose-Hubbard model on the triangular lattice with  onsite ($U$) and nearest-neighbor ($V$) repulsion and {\it positive} nearest-neighbor hopping amplitude $t$, given by the Hamiltonian
\be\label{eq:Ham}
\hat{H} &=& \sum_{\langle i, j \rangle }\left[ t  \hat{b}^\dagger_i \hat{b}_j + \text{h.c.} + V\delta \hat{n}_i\delta \hat{n}_j\right] + \frac{U}{2}\sum_i(\delta \hat{n}_i )^2 
\ee
where the first sum is over nearest-neighbor bonds $\langle i,j\rangle$, $\delta\hat{n}_i \equiv \hat{b}^\dagger_i \hat{b}_i - n_0$  and $t, U, V > 0$. Throughout, we work at fixed integer filling $n_0$ and assume $V<U$.
 We will discuss the experimental relevance of (\ref{eq:Ham}) presently, but for now we focus on its zero-temperature phase diagram. We consider three distinct phases: (i) the chiral superfluid ($\chi$SF) that breaks $U(1)$ phase, time reversal ($\mathcal{T}$) and parity ($\mathcal{P}$) symmetries \footnote{On the triangular lattice, breaking $\mathcal{P}$ and $C_6$ are equivalent.}; (ii) the chiral Mott insulator ($\chi$MI), that breaks $\mathcal{T}$ and $\mathcal{P}$; 
  and finally (iii) the trivial Mott insulator (MI), that breaks no symmetries \footnote{For $V>U$, other phases such as charge-density waves and supersolids may arise; as this regime is unlikely to be experimentally relevant, we will be silent on these.}. 
 As we will show, a {\it trivial} SF breaking only $U(1)$
 is always unstable to the $\chi$SF and is therefore irrelevant.

 The trivial insulator is simplest to describe, since it emerges even when $t\rightarrow 0$. A good variational ground state is then
\be\label{eq:trivMI}
\ket{\text{MI}} = \prod_{i} ({n_0!})^{-1/2}{(b^\dagger_i)^{n_0}}\ket{0}.
\ee
As the $\chi$SF and $\chi$MI spontaneously break $\mathcal{T}$ and $\mathcal{P}$, and emerge upon including fluctuations about the MI that are induced by the nearest-neighbor hopping, we may study them by examining the instability of the MI to developing chiral order, while breaking or retaining $U(1)$ symmetry. 

In writing variational {\it ansatze} for these phases, it is instructive to first consider the hopping term in Eq. (\ref{eq:Ham}) in momentum space.  Simple trial wavefunctions for both the chiral phases can then be written down quite straightforwardly by building suitable correlations on (\ref{eq:Ham}).  Since $t>0$, we have an `inverted' dispersion with two inequivalent minima at the Brillouin zone corners $\bK, \bK'$ (Fig.~\ref{fig:varphasediag}b).
The $\chi$SF emerges from the trivial MI upon condensing into the single-particle state at $\bK$ (a similar wavefunction exists for $\bK' = -\bK$),
\be\label{eq:chiSF}
\ket{\chi_{\text{SF}}} &=&\mathcal{N}_{SF}^{-1/2}\exp\left\{\psi \frac{\hat{b}^\dagger_{\bK}}{\sqrt{n_0+1}} + \psi^* \frac{\hat{b}_{\bK}}{\sqrt{n_0}} \right\}\ket{\text{MI}}.
\ee
On the other hand the $\chi$MI is built by dressing the trivial MI with bond correlations that take advantage of the 
positive hopping to lower the energy, by developing a non-zero current around each plaquette. The trial wavefunction condenses a particle-hole pair of the MI with the appropriate symmetry,
\be\label{eq:chiMI}
\ket{\chi_{\text{MI}}} &=&\mathcal{N}_{MI}^{-1/2} \exp\left\{\Delta\sum_{\bk} \varphi_{\bk} \hat{b}^\dagger_{\bk}\hat{b}_{\bk}\right\}\ket{\text{MI}}.\ee
where  $\varphi_{\bk} = -\varphi_{-\bk}$ parametrizes the chiral symmetry-breaking. A natural choice consistent with this symmetry is obtained by considering imaginary hopping around the triangles, which yields $\varphi_{\bk}^{(0)}= 2\sum_{i=1}^3 \sin (\bk\cdot\ba_i)$ where the $\ba_i$ are primitive vectors of the triangular lattice (Fig.~\ref{fig:varphasediag}a); intuitively, this is the simplest current pattern that can lower the energy of the MI by breaking $\mathcal{T},\mathcal{P}$ symmetries. $\mathcal{N}_{SF}, \mathcal{N}_{MI}$ are normalization constants that depend on variational parameters $\psi, \Delta$.

We now evaluate the energy of the variational {\it ansatze}  (\ref{eq:chiSF}) and (\ref{eq:chiMI}) using Hamiltonian (\ref{eq:Ham}). This is straightforward but tedious
~\cite{SuppMat}. In both cases, we work to quadratic order in the variational parameters; to this order we can ignore the dependence of the normalization constants on  $\Delta, \psi$.  
The energy per particle in the $\chi$SF state is 
\be
\mathcal{E}_{\chi_{\text{SF}}}^{(\psi)} 
\label{eq:chiSFE}
&=& |\psi|^2\left[\xi_{\bK} (\sqrt{n_0+1} +\sqrt{n_0})^2 +  U  \right] + O(|\psi|^4)
\ee
where $\xi_{\bK} = -3t$ is the single-particle kinetic energy at the band minimum. Note that a condensate at any arbitrary momentum away from the minima at $\pm\bK$ will have a greater energy, including the trivial SF at $\bk=0$. We are therefore justified in our neglect of such possibilities at the outset. Observe that $\mathcal{E}_{\chi_{\text{SF}}}$ is independent of $V$ to this order; this will prove important in the competition with the $\chi$MI.

Turning to the $\chi$MI, we find for its variational energy
\be
\mathcal{E}_{\chi_{\text{MI}}}^{(\Delta)} 
&=& 3\Delta^2 n_0(n_0+1)\left[U-V-4t(2n_0+1)  \right]
+ O(\Delta^4)\,\,\,\,\,\,\label{eq:chiMIE}
\ee
which depends on the nearest-neighbor repulsion. From (\ref{eq:chiSFE}) and (\ref{eq:chiMIE}), it is evident that within the variational approach, the trivial MI gives way to the $\chi$SF for $U < U_c = 3t  (\sqrt{n_0+1} +\sqrt{n_0})^2$, while it is unstable to the $\chi$MI for $U< 4t(2n_0+1) +V$. Thus, for any $n_0$ and at $V=0$, at mean-field level the instability of the MI to $\chi$SF order preempts that to the $\chi$MI. However, for nonzero $V$, the $\chi$MI lowers its energy relative to the $\chi$SF, and emerges as a stable phase if 
 \be
U> U_c,\quad U - \frac{4(2n_0+1)}{3  (\sqrt{n_0+1} +\sqrt{n_0})^2}U_c < V<U,
\ee
 where in the final expression we have eliminated $t$ in favor of $U_c$. The resulting phase diagram 
 is sketched in Fig.~\ref{fig:varphasediag}c. 

The different states in the phase diagram should have clear signatures in the momentum distribution function that can be obtained from time-of-flight measurements. As shown in Fig.~\ref{fig:varphasediag}d,
the $\chi$SF  is characterized by a sharp peak at the condensate momenta $\bK$ or $\bK'$. In the $\chi$MI these peaks broaden, but the parity asymmetry, reflecting broken $\mathcal{T},\mathcal{P}$ and $C_6$ symmetries, persists. Full symmetry is restored in the MI. The images  in Fig.~\ref{fig:varphasediag}d are computed using the variational {\it ansatze}, for which the deviation of the momentum distribution of the $\chi$MI from the symmetric distribution of the MI is given by
$\langle \hat{n}_{\bk}\rangle_{\chi_{\text{MI}}}  - \langle \hat{n}_{\bk}\rangle_{\text{MI}}  = 2n_0 (n_0+1)\varphi_{\bk}\Delta$ in lowest order in $\Delta$ \cite{SuppMat}. 
Note that within  this variational treatment the $\chi$MI requires a minimal value of the nearest neighbor repulsion $V$ (although it appears that the $\chi$MI persists at $V=0$ on ladders, see below). Fortunately, the metastable upper band condensates that give rise to complex superfluids also have an intrinsic mechanism that can generate substantial and tunable nearest-neighbor interactions. The required positive hopping matrix element in such lattices stems from hopping through a lower energy intermediate site. The same intermediate site also mediates a nearest neighbor interaction~\cite{SuppMat}.  

\begin{figure}
\includegraphics[width=0.9\columnwidth]{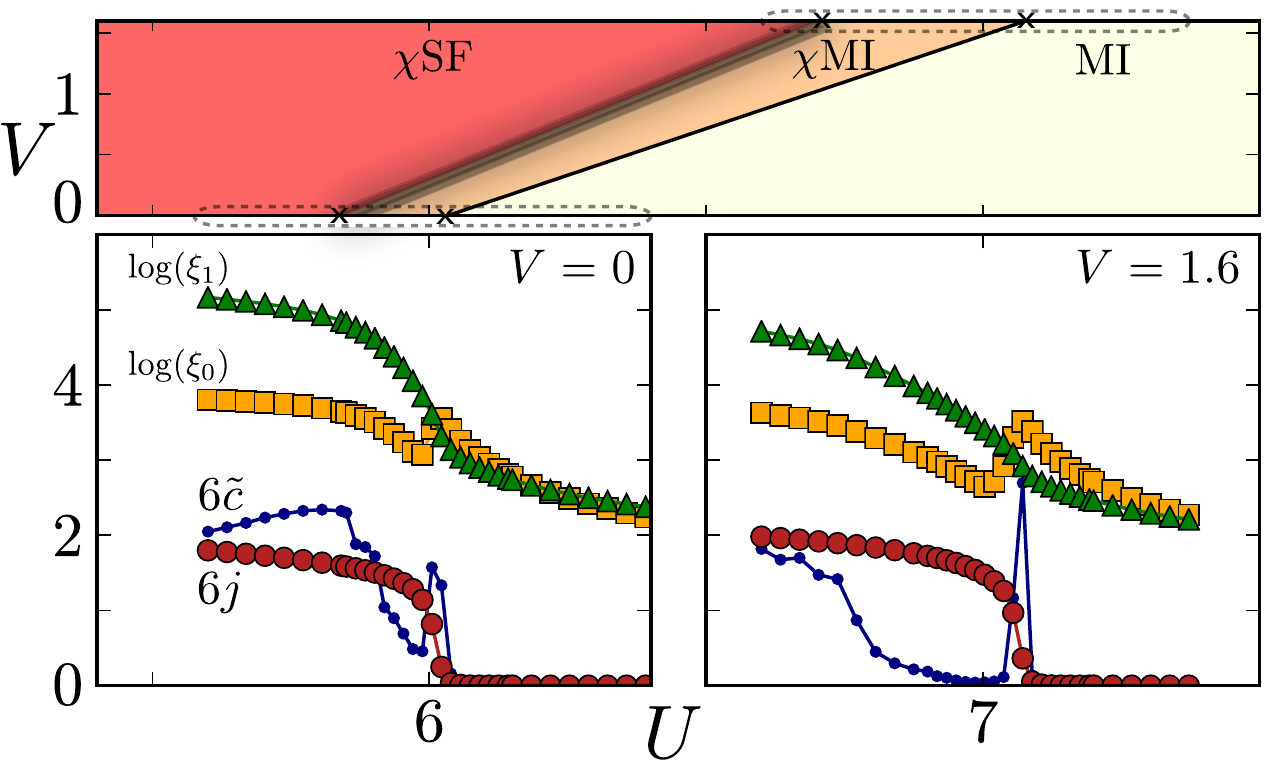}
\caption{{\bf \label{fig:4LLfig} iDMRG on 4-leg ladder.} Varying $U$ for $V=0, 1.6$  keeping 1600 Schmidt states (dotted), we plot neutral/charge correlation lengths $\xi_{0,1}$ (from $\langle \hat{n}_i \hat{n}_0\rangle,\langle \hat{b}^\dagger_i\hat{b}_0\rangle$), bond current $j$ ($\chi$MI order parameter) and logarithmic derivative of entanglement entropy with bond dimension ($\tilde{c}$, finite only for gapless phases). At MI$\rightarrow\chi$MI transition $\xi_0$ and $\tilde{c}$ jump and $j$ onsets, indicating Ising ordering. In the $\chi$MI, $\tilde{c}=0, j\neq0$ indicating gap and broken symmetry; at the BKT transition and in $\chi$SF, $\tilde{c}\neq0$.  FES is biased towards the $\chi$SF, so {\it underestimates} the regime of the $\chi$MI, indicated by shaded BKT~line.~\cite{SuppMat}}
\end{figure}

{\it Bosonization and iDMRG on Ladders.---} 
We now extend our mean-field results by studying one-dimensional analogs of our original problem: frustrated triangular ladders. As a first step, we establish the stability of the phases suggested in the mean-field picture on an infinitely long, 4-leg ladder with $U,V$ and isotropic couplings using the iDMRG algorithm. 
We work at $n_0 = 1$ and truncate the Hilbert space to at most two bosons per site.
While iDMRG has no finite size effects, for critical states the finite number of Schmidt states kept, $\chi$, cuts off the correlations at a `finite entanglement' length scale $\xi_{FE} \sim \chi^\kappa$~\cite{pollmann2009}.
We can therefore extract critical properties by replacing finite-size scaling with finite entanglement scaling (FES),  where  $\xi_{FE}$ plays a role analogous to the system size~\cite{tagliacozzo2008, Kjall2013}.
In Fig.~\ref{fig:4LLfig} we present results establishing the existence of all three phases for $V=0, 1.6 $. While definitively concluding that the $\chi$MI persists to $V=0$ in $d=2$ requires a scaling analysis beyond our present reach, these results suggest that relatively small $V$ may suffice to stabilize the $\chi$MI once fluctuations beyond mean-field are included.

We gain more insight by studying a {\it two}-leg ladder, where we can perform iDMRG simulations of sufficient precision to extract critical behavior. An attendant bosonization analysis elucidates the long-wavelength field theory governing the phase diagram. We first recast the problem into a slightly different form: we set $V=0$, and  -- anticipating that they will renormalize independently under RG flow --
 we allow the `leg' ($t$) and  `rung'($t_\perp$) couplings to differ at the outset:
\begin{eqnarray}
\label{eq:1dH}
\hat{H}  = {t}\sum_{\langle i, j\rangle} b^\dagger_i b_j + {t_\perp}\sum_{\langle \langle i, j \rangle \rangle} b^\dagger_i b_j  +\text{h.c.} + \frac{1}{2} \sum_i (\delta\hat{n}_i)^2
\end{eqnarray}
where the sums run over nearest neighbor bonds on legs and rungs, and we have set $U=1$ for simplicity.
The hopping is frustrated (the dispersion has two minima) for $t >  t_\perp/4 > 0$. 

\begin{figure}
\includegraphics{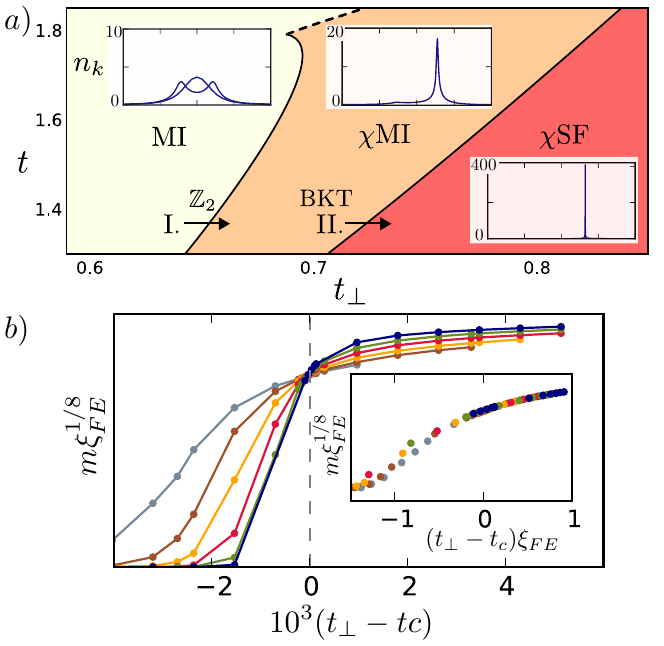}
\caption{\textbf{iDMRG on two-leg Ladder.} (a) Phase diagram for $t,t_\perp>0$ showing all three phases and cuts along which critical scaling is studied, with $\langle n_{\bk}\rangle$ inset. Note that in the MI $\langle n_{\bk}\rangle$ can be bimodal close to the $\chi$MI; also, the MI-$\chi$MI phase boundary is first-order along the dashed segment. (b) Finite Entanglement Scaling (FES) at MI $\to \chi$MI transition, with $m \equiv \Im \langle b^\dagger_{i} b_{i+1} \rangle$. For each 
 $t_\perp$, we obtain several increasingly accurate ground states 
 by increasing iDMRG bond dimension $\chi$;  
their correlation length  $\xi_{FE}$ 
serves as the FES parameter. 
At the Ising transition, $m$  and $t_\perp- t_c$  have mass dimension $1/8$ and $1$.
Plotting $m \xi_{FE}^{1/8}$, the FES curves  
cross at $t_c$. (Inset)
Plotting $m \xi_{FE}^{1/8}$ as a function of $(t - t_c) \xi_{FE}$, the data collapses.  
}
\label{fig:1d_phases}
\end{figure}

The full phase diagram for $t, t_\perp > 0$ is quite rich and contains chiral and non-chiral SFs (with algebraic order), MI, and $\chi$MI phases, as well as an interesting tricritical Ising point. 
We focus on the range of $t, t_\perp$ most relevant to our 2D interests (Fig. \ref{fig:1d_phases}), 
and show (i) that a $\chi$MI separates a $\chi$SF and a Mott insulator, and (ii) that the transitions between them are of BKT and Ising types respectively.

Before bosonizing~\cite{giamarchi2004quantum}, we pass to a different representation, defining rotors $\theta^\alpha$ on the two legs ($\alpha = 1, 2$). In coordinates  \footnote{This notation makes the gradient expansion particularly simple.} where the sites of leg $\alpha$ lie at $\mathbb{Z} + e_\alpha$, with $e_{1,2} = \pm 1/4$ and for $n_0\gg 1$ we can approximate Eq. \eqref{eq:1dH} by a rotor Hamiltonian $\hat{H}_1 =\sum_x \hat{\mathcal{H}}_x$, where
\be\label{eq:1Drotors}
 \hat{\mathcal{H}}_x &=& 
 \sum_{\alpha=1}^2\left\{-J\cos\left(\hat{\theta}^\alpha_{x+e_\alpha+1}\!-\!\hat{\theta}^\alpha_{x+e_\alpha}\right) +\frac{1}{2 J} \left(\hat{n}^{\alpha}_{x+e_\alpha}\right)^2\right\} \\& & 
  -J_{\perp}\left\{\cos\left(\hat{\theta}^1_{x+e_1} \!-\!\hat{\theta}^2_{x+e_2}\right)\!-\!\cos\left(\hat{\theta}^1_{x+e_1} \!-\!\hat{\theta}^2_{x+e_2+1}\right)\right\},\nonumber
\ee
with the usual canonical commutator, $[\hat{n}^{\alpha}_x, \hat{\theta}^{\beta}_{x'}] =i\delta_{\alpha\beta}\delta_{x,x'}$.
$J, J_\perp$ are phenomenological constants roughly set by $t, t_\perp$.

We now take the  continuum limit of $\hat{H}_1$, perform a gradient expansion, and  introduce conjugate fields $\phi^\alpha(x)$ satisfying $[{\theta}^\alpha(x), \pi^{-1}\partial_x \phi^\beta(x')] = i\delta(x-x')\delta_{\alpha\beta}$. It is convenient to canonically change variables to the symmetric and antisymmetric combinations $\theta_{+} \equiv \frac{1}{2}(\theta^{1} + \theta^2), \theta_-\equiv \theta^1-\theta^2$,   $\phi_+ \equiv \phi^1+\phi^2, \phi_- \equiv\frac{1}{2}(\phi^1-\phi^2)$. After  some algebra,
we find  $\hat{H_1} \approx \int dx\mathcal{H} = \int dx(\hat{\mathcal{H}}_0+\hat{\mathcal{H}}_{\perp})$, where 
\be\label{eq:1dbosform}
\hat{\mathcal{H}}_{0} &=& \frac{1}{4 J} \left(\frac{\partial_x\phi_+}{\pi}\right)^2 \!+\! J(\partial_x\theta_+)^2
\!+\!\frac{1}{J} \left(\frac{\partial_x\phi_-}{\pi}\right)^2\!+\!  \frac{J}{4}(\partial_x\theta_-)^2  \nonumber\\
\hat{\mathcal{H}}_{\perp} &=& J_\perp \sin\theta_- \partial_x\theta_+ +  \zeta_c\cos(\phi_+)\cos(2\phi_-)\ee
and  $\zeta_c$ is the  fugacity of kinks (phase slips) that restore the phase periodicity that is lost when performing the gradient expansion. The elementary phase slip 
is a composite of a $\pi$ slip in $\theta_+$ and a $2\pi$ slip in $\theta_-$, which is a rewriting of $2\pi$ phase-slips in $\theta^{1,2}$ into the new variables $\theta_{\pm}$. 

The coupled sine-Gordon theory (\ref{eq:1dbosform}) suffices to capture the different phases and transitions between them.
In the MI both $\theta_-$ and $\theta_+$ are disordered.
In the $\chi$MI,  $\theta_-$ is locked while $\theta_+$ is disordered.
In the $\chi$SF, $\theta_-$ is locked, while $\theta_+$ is algebraically ordered, with $\langle \partial_x \theta_+ \rangle \neq 0$. The coupling between the orders, controlled by $J_\perp$, is unusual: the supercurrent $\partial_x\theta_+$ couples to the $\mathcal{T},\mathcal{P}$ breaking Ising variable $\sin\theta_-$.

To study the MI-$\chi$MI transition we use the fact that  $\phi_+$ is locked in the Mott phases to write $\zeta_c\cos(\phi_+)\cos(2\phi_-) \sim \zeta_{\text{eff}} \cos(2\phi_-)$.
Integrating out the gapped $\theta_+$ fluctuations from  (\ref{eq:1dbosform}) yields a term $\cos(2\theta_-)$ at order $ \mathcal{O}(\frac{J_\perp^2}{8J})$, and (\ref{eq:1dbosform}) reduces to a double sine-Gordon model for $\theta_-$,
\be
\mathcal{H}&\approx& \frac{(\partial_x\phi_-)^2}{\pi^2J} + \frac{J}{4}(\partial_x\theta_-)^2
 +\zeta_{\text{eff}} \cos(2\phi_-) +  \frac{J_\perp^2}{8J}\cos(2\theta_-)\nonumber
\ee
 where the two cosines compete to stabilize the MI and $\chi$MI respectively. For sufficiently large $J_\perp$, the $\cos(2\theta_-)$ term will lock $\theta_- = \pm\pi/2$. In the resulting phase $\langle \sin \theta_-\rangle \neq 0$, while $\theta_+$ remains disordered. The current coupling in $\hat{\mathcal{H}}_\perp$ implies $\langle \partial_x \theta_+\rangle \neq 0$, which corresponds to the chiral order. The critical point can be studied via refermionization~\cite{giamarchi2004quantum} and is described by a single gapless Majorana fermion, with central charge $c=1/2$ --- {\it i.e.}, it is a standard $\mathbb{Z}_2$ Ising transition.

For the $\chi$MI-$\chi$SF transition, we may integrate out $\theta_-$ as it is locked in both phases, 
yielding a standard BKT transition driven by vortices created by $\cos2\phi_+$, which is generated from $\mathcal{H}_\perp$.
When these vortices bind, $\theta_+$ develops algebraic order, but the minimum of $\partial_x\theta_+$ is shifted to a finite $K_0 = \frac{J_\perp}{2J}\langle\sin(\theta_-)\rangle$ determined by the chiral symmetry-breaking.
As a consequence, the large $J$  superfluid has nonzero current around triangular plaquettes, which breaks chiral symmetry. 

At $J_\perp =0$
 the two legs decouple, and form independent superfluids for large $J$.  
Along the decoupled line $c=2$ and the scaling dimension of the coupling is $[\sin\theta_-\partial_x\theta_+] = 1 + (8\pi J)^{-1}$, which is always relevant in the critical phase. Hence for infinitesimal $J_\perp$ the system will flow to the $\chi$SF.
While it is likely that the $\chi$MI also persists to $J_\perp=0$ so that all three phases meet at a triple point, we defer this to future work \cite{ZPRA-future}.

We confirm this picture numerically, by simulating the Hamiltonian of Eq.~\eqref{eq:1dH} using iDMRG. 
The relevant part of the 1D phase diagram (Fig.~\ref{fig:1d_phases}a) clearly shows the  MI, $\chi$MI and $\chi$SF, distinguished  by their 1D momentum distributions, shown inset.  For $t\gtrsim 1.8$, the MI$\rightarrow$$\chi$MI line has a first-order segment, a possible sign of tricritical behavior \cite{ZPRA-future}.  
To determine the universality class of the continuous transitions, we perform FES along 
two cuts. 
From cut I, we confirm that the MI$\rightarrow$$\chi$MI transition is Ising, with 
the current on a rung, $m \equiv \Im b^\dagger_{i} b_{i+1}$ serving as the order parameter, and $t_\perp$ as coupling constant.
The FES collapse of the order parameter  $m(t, \xi_{FE}) = \xi_{FE}^{-1/8} \Phi[ \xi_{FE}(t - t_c)]$ (Fig.~\ref{fig:1d_phases}b) is consistent with the Ising scaling dimensions; in addition, though not shown, there is clear FES evidence that $c=\frac{1}{2}$. From cut II we conclude that the $\chi$MI $\rightarrow$$\chi$SF transition is BKT, 
with a  universal jump in the superfluid stiffness.

{\it Concluding Remarks.---}
We have shown that the transition from the $\mathcal{T}$-breaking superfluid, observed in recent experiments on 2D optical lattices~\cite{Wirth2010,Olschlager2011}, to a trivial MI upon increasing optical lattice depth can occur via an intermediate 
 chiral MI phase. Using variational wavefunctions we mapped the phase diagram of a specific model of interacting bosons on the frustrated triangular lattice.  We recover the same three phases ($\chi$SF, $\chi$MI and MI) using bosonization and iDMRG of finite-width ladders. 
Unconventional Bose insulators can thus be observed and investigated in current experiments.

In closing, we note that the $\chi$MI phase bears a certain family resemblance to $\mathcal{T}$-breaking phases that have been proposed to emerge from strong electronic correlations in high-$T_c$ superconductors \cite{NayakWilczekDDW,VarmaNFLPairing,VarmaPseudoGap,NayakDensityWave,ChakravartyCuprates}. The present work suggests it may be possible to realize similar phases in purely bosonic models.

 {\it Acknowledgements. --- }We thank D.M. Stamper-Kurn for discussions stimulating this work, D. V\'arjas, J.E. Moore, and C.K. Thomas for comments on the manuscript, and NSF GRFP Grant DGE 1106400 (MPZ), the Simons Foundation (SAP), the Swiss National Science Foundation (AR), the ISF, BSF, ERC Synergy UQUAM program and the Miller Institute at UC Berkeley (EA) and NSF Grant No. 1066293  at the Aspen Center for Physics (SAP, EA) for support.
\bibliography{ChiralMott-bib}

\begin{thebibliography}{26}%
\makeatletter
\providecommand \@ifxundefined [1]{%
 \@ifx{#1\undefined}
}%
\providecommand \@ifnum [1]{%
 \ifnum #1\expandafter \@firstoftwo
 \else \expandafter \@secondoftwo
 \fi
}%
\providecommand \@ifx [1]{%
 \ifx #1\expandafter \@firstoftwo
 \else \expandafter \@secondoftwo
 \fi
}%
\providecommand \natexlab [1]{#1}%
\providecommand \enquote  [1]{``#1''}%
\providecommand \bibnamefont  [1]{#1}%
\providecommand \bibfnamefont [1]{#1}%
\providecommand \citenamefont [1]{#1}%
\providecommand \href@noop [0]{\@secondoftwo}%
\providecommand \href [0]{\begingroup \@sanitize@url \@href}%
\providecommand \@href[1]{\@@startlink{#1}\@@href}%
\providecommand \@@href[1]{\endgroup#1\@@endlink}%
\providecommand \@sanitize@url [0]{\catcode `\\12\catcode `\$12\catcode
  `\&12\catcode `\#12\catcode `\^12\catcode `\_12\catcode `\%12\relax}%
\providecommand \@@startlink[1]{}%
\providecommand \@@endlink[0]{}%
\providecommand \url  [0]{\begingroup\@sanitize@url \@url }%
\providecommand \@url [1]{\endgroup\@href {#1}{\urlprefix }}%
\providecommand \urlprefix  [0]{URL }%
\providecommand \Eprint [0]{\href }%
\providecommand \doibase [0]{http://dx.doi.org/}%
\providecommand \selectlanguage [0]{\@gobble}%
\providecommand \bibinfo  [0]{\@secondoftwo}%
\providecommand \bibfield  [0]{\@secondoftwo}%
\providecommand \translation [1]{[#1]}%
\providecommand \BibitemOpen [0]{}%
\providecommand \bibitemStop [0]{}%
\providecommand \bibitemNoStop [0]{.\EOS\space}%
\providecommand \EOS [0]{\spacefactor3000\relax}%
\providecommand \BibitemShut  [1]{\csname bibitem#1\endcsname}%
\let\auto@bib@innerbib\@empty
\bibitem [{\citenamefont {Greiner}\ \emph {et~al.}(2002)\citenamefont
  {Greiner}, \citenamefont {Mandel}, \citenamefont {Esslinger}, \citenamefont
  {Hansch},\ and\ \citenamefont {Bloch}}]{Greiner2002}%
  \BibitemOpen
  \bibfield  {author} {\bibinfo {author} {\bibfnamefont {M.}~\bibnamefont
  {Greiner}}, \bibinfo {author} {\bibfnamefont {O.}~\bibnamefont {Mandel}},
  \bibinfo {author} {\bibfnamefont {T.}~\bibnamefont {Esslinger}}, \bibinfo
  {author} {\bibfnamefont {T.~W.}\ \bibnamefont {Hansch}}, \ and\ \bibinfo
  {author} {\bibfnamefont {I.}~\bibnamefont {Bloch}},\ }\href {\doibase
  10.1038/415039a} {\bibfield  {journal} {\bibinfo  {journal} {Nature}\
  }\textbf {\bibinfo {volume} {415}},\ \bibinfo {pages} {39} (\bibinfo {year}
  {2002})}\BibitemShut {NoStop}%
\bibitem [{\citenamefont {Parameswaran}\ \emph {et~al.}(2013)\citenamefont
  {Parameswaran} \emph {et~al.}}]{KagomeWannier}%
  \BibitemOpen
  \bibfield  {author} {\bibinfo {author} {\bibfnamefont {S.~A.}\ \bibnamefont
  {Parameswaran}} \emph {et~al.},\ }\href {\doibase
  10.1103/PhysRevLett.110.125301} {\bibfield  {journal} {\bibinfo  {journal}
  {Phys. Rev. Lett.}\ }\textbf {\bibinfo {volume} {110}},\ \bibinfo {pages}
  {125301} (\bibinfo {year} {2013})}\BibitemShut {NoStop}%
\bibitem [{\citenamefont {Kimchi}\ \emph {et~al.}(2012)\citenamefont {Kimchi}
  \emph {et~al.}}]{HoneycombVoronoi}%
  \BibitemOpen
  \bibfield  {author} {\bibinfo {author} {\bibfnamefont {I.}~\bibnamefont
  {Kimchi}} \emph {et~al.},\ }\href@noop {} {\  (\bibinfo {year} {2012})},\
  \Eprint {http://arxiv.org/abs/1207.0498} {arXiv:1207.0498} \BibitemShut
  {NoStop}%
\bibitem [{\citenamefont {Wirth}\ \emph {et~al.}(2010)\citenamefont {Wirth},
  \citenamefont {\"{O}lschl\"{a}ger},\ and\ \citenamefont
  {Hemmerich}}]{Wirth2010}%
  \BibitemOpen
  \bibfield  {author} {\bibinfo {author} {\bibfnamefont {G.}~\bibnamefont
  {Wirth}}, \bibinfo {author} {\bibfnamefont {M.}~\bibnamefont
  {\"{O}lschl\"{a}ger}}, \ and\ \bibinfo {author} {\bibfnamefont
  {A.}~\bibnamefont {Hemmerich}},\ }\href {\doibase 10.1038/nphys1857}
  {\bibfield  {journal} {\bibinfo  {journal} {Nature Physics}\ }\textbf
  {\bibinfo {volume} {7}},\ \bibinfo {pages} {147} (\bibinfo {year}
  {2010})}\BibitemShut {NoStop}%
\bibitem [{\citenamefont {\"{O}lschl\"{a}ger}\ \emph
  {et~al.}(2011)\citenamefont {\"{O}lschl\"{a}ger}, \citenamefont {Wirth},\
  and\ \citenamefont {Hemmerich}}]{Olschlager2011}%
  \BibitemOpen
  \bibfield  {author} {\bibinfo {author} {\bibfnamefont {M.}~\bibnamefont
  {\"{O}lschl\"{a}ger}}, \bibinfo {author} {\bibfnamefont {G.}~\bibnamefont
  {Wirth}}, \ and\ \bibinfo {author} {\bibfnamefont {A.}~\bibnamefont
  {Hemmerich}},\ }\href {\doibase 10.1103/PhysRevLett.106.015302} {\bibfield
  {journal} {\bibinfo  {journal} {Physical Review Letters}\ }\textbf {\bibinfo
  {volume} {106}},\ \bibinfo {pages} {015302} (\bibinfo {year}
  {2011})}\BibitemShut {NoStop}%
\bibitem [{\citenamefont {Struck}\ \emph {et~al.}(2011)\citenamefont {Struck},
  \citenamefont {\"{O}lschl\"{a}ger}, \citenamefont {{Le Targat}},
  \citenamefont {Soltan-Panahi}, \citenamefont {Eckardt}, \citenamefont
  {Lewenstein}, \citenamefont {Windpassinger},\ and\ \citenamefont
  {Sengstock}}]{Struck2011}%
  \BibitemOpen
  \bibfield  {author} {\bibinfo {author} {\bibfnamefont {J.}~\bibnamefont
  {Struck}}, \bibinfo {author} {\bibfnamefont {C.}~\bibnamefont
  {\"{O}lschl\"{a}ger}}, \bibinfo {author} {\bibfnamefont {R.}~\bibnamefont
  {{Le Targat}}}, \bibinfo {author} {\bibfnamefont {P.}~\bibnamefont
  {Soltan-Panahi}}, \bibinfo {author} {\bibfnamefont {A.}~\bibnamefont
  {Eckardt}}, \bibinfo {author} {\bibfnamefont {M.}~\bibnamefont {Lewenstein}},
  \bibinfo {author} {\bibfnamefont {P.}~\bibnamefont {Windpassinger}}, \ and\
  \bibinfo {author} {\bibfnamefont {K.}~\bibnamefont {Sengstock}},\ }\href
  {\doibase 10.1126/science.1207239} {\bibfield  {journal} {\bibinfo  {journal}
  {Science (New York, N.Y.)}\ }\textbf {\bibinfo {volume} {333}},\ \bibinfo
  {pages} {996} (\bibinfo {year} {2011})}\BibitemShut {NoStop}%
\bibitem [{\citenamefont {Lee}\ \emph {et~al.}(1984)\citenamefont {Lee},
  \citenamefont {Joannopoulos}, \citenamefont {Negele},\ and\ \citenamefont
  {Landau}}]{Lee1984}%
  \BibitemOpen
  \bibfield  {author} {\bibinfo {author} {\bibfnamefont {D.-H.}\ \bibnamefont
  {Lee}}, \bibinfo {author} {\bibfnamefont {J.~D.}\ \bibnamefont
  {Joannopoulos}}, \bibinfo {author} {\bibfnamefont {J.~W.}\ \bibnamefont
  {Negele}}, \ and\ \bibinfo {author} {\bibfnamefont {D.~P.}\ \bibnamefont
  {Landau}},\ }\href {\doibase 10.1103/PhysRevLett.52.433} {\bibfield
  {journal} {\bibinfo  {journal} {Physical Review Letters}\ }\textbf {\bibinfo
  {volume} {52}},\ \bibinfo {pages} {433} (\bibinfo {year} {1984})}\BibitemShut
  {NoStop}%
\bibitem [{\citenamefont {Dhar}\ \emph {et~al.}(2012)\citenamefont {Dhar},
  \citenamefont {Maji}, \citenamefont {Mishra}, \citenamefont {Pai},
  \citenamefont {Mukerjee},\ and\ \citenamefont {Paramekanti}}]{Dhar2012}%
  \BibitemOpen
  \bibfield  {author} {\bibinfo {author} {\bibfnamefont {A.}~\bibnamefont
  {Dhar}}, \bibinfo {author} {\bibfnamefont {M.}~\bibnamefont {Maji}}, \bibinfo
  {author} {\bibfnamefont {T.}~\bibnamefont {Mishra}}, \bibinfo {author}
  {\bibfnamefont {R.~V.}\ \bibnamefont {Pai}}, \bibinfo {author} {\bibfnamefont
  {S.}~\bibnamefont {Mukerjee}}, \ and\ \bibinfo {author} {\bibfnamefont
  {A.}~\bibnamefont {Paramekanti}},\ }\href {\doibase
  10.1103/PhysRevA.85.041602} {\bibfield  {journal} {\bibinfo  {journal}
  {Physical Review A}\ }\textbf {\bibinfo {volume} {85}},\ \bibinfo {pages}
  {041602} (\bibinfo {year} {2012})}\BibitemShut {NoStop}%
\bibitem [{\citenamefont {Dhar}\ \emph {et~al.}(2013)\citenamefont {Dhar},
  \citenamefont {Mishra}, \citenamefont {Maji}, \citenamefont {Pai},
  \citenamefont {Mukerjee},\ and\ \citenamefont {Paramekanti}}]{Dhar2013}%
  \BibitemOpen
  \bibfield  {author} {\bibinfo {author} {\bibfnamefont {A.}~\bibnamefont
  {Dhar}}, \bibinfo {author} {\bibfnamefont {T.}~\bibnamefont {Mishra}},
  \bibinfo {author} {\bibfnamefont {M.}~\bibnamefont {Maji}}, \bibinfo {author}
  {\bibfnamefont {R.~V.}\ \bibnamefont {Pai}}, \bibinfo {author} {\bibfnamefont
  {S.}~\bibnamefont {Mukerjee}}, \ and\ \bibinfo {author} {\bibfnamefont
  {A.}~\bibnamefont {Paramekanti}},\ }\href {\doibase
  10.1103/PhysRevB.87.174501} {\bibfield  {journal} {\bibinfo  {journal}
  {Physical Review B}\ }\textbf {\bibinfo {volume} {87}},\ \bibinfo {pages}
  {174501} (\bibinfo {year} {2013})}\BibitemShut {NoStop}%
\bibitem [{\citenamefont {Chubukov}\ and\ \citenamefont
  {Starykh}(2013)}]{Chubukov:2013}%
  \BibitemOpen
  \bibfield  {author} {\bibinfo {author} {\bibfnamefont {A.~V.}\ \bibnamefont
  {Chubukov}}\ and\ \bibinfo {author} {\bibfnamefont {O.~A.}\ \bibnamefont
  {Starykh}},\ }\href {\doibase 10.1103/PhysRevLett.110.217210} {\bibfield
  {journal} {\bibinfo  {journal} {Physical Review Letters}\ }\textbf {\bibinfo
  {volume} {110}},\ \bibinfo {pages} {217210} (\bibinfo {year}
  {2013})}\BibitemShut {NoStop}%
\bibitem [{\citenamefont {White}(1992)}]{White-1992}%
  \BibitemOpen
  \bibfield  {author} {\bibinfo {author} {\bibfnamefont {S.~R.}\ \bibnamefont
  {White}},\ }\href {\doibase 10.1103/PhysRevLett.69.2863} {\bibfield
  {journal} {\bibinfo  {journal} {Phys. Rev. Lett.}\ }\textbf {\bibinfo
  {volume} {69}},\ \bibinfo {pages} {2863} (\bibinfo {year}
  {1992})}\BibitemShut {NoStop}%
\bibitem [{\citenamefont {McCulloch}(2008)}]{McCulloch-2008}%
  \BibitemOpen
  \bibfield  {author} {\bibinfo {author} {\bibfnamefont {I.~P.}\ \bibnamefont
  {McCulloch}},\ }\href@noop {} {\enquote {\bibinfo {title} {Infinite size
  density matrix renormalization group, revisited},}\ } (\bibinfo {year}
  {2008}),\ \bibinfo {note} {unpublished},\ \Eprint
  {http://arxiv.org/abs/0804.2509} {arXiv:0804.2509} \BibitemShut {NoStop}%
\bibitem [{\citenamefont {Kj\"all}\ \emph {et~al.}(2013)\citenamefont
  {Kj\"all}, \citenamefont {Zaletel}, \citenamefont {Mong}, \citenamefont
  {Bardarson},\ and\ \citenamefont {Pollmann}}]{Kjall2013}%
  \BibitemOpen
  \bibfield  {author} {\bibinfo {author} {\bibfnamefont {J.~A.}\ \bibnamefont
  {Kj\"all}}, \bibinfo {author} {\bibfnamefont {M.~P.}\ \bibnamefont
  {Zaletel}}, \bibinfo {author} {\bibfnamefont {R.~S.~K.}\ \bibnamefont
  {Mong}}, \bibinfo {author} {\bibfnamefont {J.~H.}\ \bibnamefont {Bardarson}},
  \ and\ \bibinfo {author} {\bibfnamefont {F.}~\bibnamefont {Pollmann}},\
  }\href {\doibase 10.1103/PhysRevB.87.235106} {\bibfield  {journal} {\bibinfo
  {journal} {Phys. Rev. B}\ }\textbf {\bibinfo {volume} {87}},\ \bibinfo
  {pages} {235106} (\bibinfo {year} {2013})}\BibitemShut {NoStop}%
\bibitem [{Note1()}]{Note1}%
  \BibitemOpen
  \bibinfo {note} {On the triangular lattice, breaking $\protect \mathcal {P}$
  and $C_6$ are equivalent.}\BibitemShut {Stop}%
\bibitem [{Note2()}]{Note2}%
  \BibitemOpen
  \bibinfo {note} {For $V>U$, other phases such as charge-density waves and
  supersolids may arise; as this regime is unlikely to be experimentally
  relevant, we will be silent on these.}\BibitemShut {Stop}%
\bibitem [{Sup()}]{SuppMat}%
  \BibitemOpen
  \href@noop {} {\enquote {\bibinfo {title} {Supplementary material},}\
  }\BibitemShut {NoStop}%
\bibitem [{\citenamefont {Pollmann}\ \emph {et~al.}(2009)\citenamefont
  {Pollmann}, \citenamefont {Mukerjee}, \citenamefont {Turner},\ and\
  \citenamefont {Moore}}]{pollmann2009}%
  \BibitemOpen
  \bibfield  {author} {\bibinfo {author} {\bibfnamefont {F.}~\bibnamefont
  {Pollmann}}, \bibinfo {author} {\bibfnamefont {S.}~\bibnamefont {Mukerjee}},
  \bibinfo {author} {\bibfnamefont {A.~M.}\ \bibnamefont {Turner}}, \ and\
  \bibinfo {author} {\bibfnamefont {J.~E.}\ \bibnamefont {Moore}},\ }\href
  {\doibase 10.1103/PhysRevLett.102.255701} {\bibfield  {journal} {\bibinfo
  {journal} {Phys. Rev. Lett.}\ }\textbf {\bibinfo {volume} {102}},\ \bibinfo
  {pages} {255701} (\bibinfo {year} {2009})}\BibitemShut {NoStop}%
\bibitem [{\citenamefont {Tagliacozzo}\ \emph {et~al.}(2008)\citenamefont
  {Tagliacozzo}, \citenamefont {de~Oliveira}, \citenamefont {Iblisdir},\ and\
  \citenamefont {Latorre}}]{tagliacozzo2008}%
  \BibitemOpen
  \bibfield  {author} {\bibinfo {author} {\bibfnamefont {L.}~\bibnamefont
  {Tagliacozzo}}, \bibinfo {author} {\bibfnamefont {T.~R.}\ \bibnamefont
  {de~Oliveira}}, \bibinfo {author} {\bibfnamefont {S.}~\bibnamefont
  {Iblisdir}}, \ and\ \bibinfo {author} {\bibfnamefont {J.~I.}\ \bibnamefont
  {Latorre}},\ }\href {\doibase 10.1103/PhysRevB.78.024410} {\bibfield
  {journal} {\bibinfo  {journal} {Phys. Rev. B}\ }\textbf {\bibinfo {volume}
  {78}},\ \bibinfo {pages} {024410} (\bibinfo {year} {2008})}\BibitemShut
  {NoStop}%
\bibitem [{\citenamefont {Giamarchi}(2004)}]{giamarchi2004quantum}%
  \BibitemOpen
  \bibfield  {author} {\bibinfo {author} {\bibfnamefont {T.}~\bibnamefont
  {Giamarchi}},\ }\href@noop {} {\emph {\bibinfo {title} {Quantum Physics in
  One Dimension}}},\ International Series of Monographs on Physics\ (\bibinfo
  {publisher} {Clarendon Press},\ \bibinfo {address} {Oxford},\ \bibinfo {year}
  {2004})\BibitemShut {NoStop}%
\bibitem [{Note3()}]{Note3}%
  \BibitemOpen
  \bibinfo {note} {This notation makes the gradient expansion particularly
  simple.}\BibitemShut {Stop}%
\bibitem [{\citenamefont {Zaletel}\ \emph {et~al.}(2013)\citenamefont
  {Zaletel}, \citenamefont {Parameswaran}, \citenamefont {R\"{u}egg},\ and\
  \citenamefont {Altman}}]{ZPRA-future}%
  \BibitemOpen
  \bibfield  {author} {\bibinfo {author} {\bibfnamefont {M.~P.}\ \bibnamefont
  {Zaletel}}, \bibinfo {author} {\bibfnamefont {S.~A.}\ \bibnamefont
  {Parameswaran}}, \bibinfo {author} {\bibfnamefont {A.}~\bibnamefont
  {R\"{u}egg}}, \ and\ \bibinfo {author} {\bibfnamefont {E.}~\bibnamefont
  {Altman}},\ }\href@noop {} {} (\bibinfo {year} {2013}),\ \bibinfo {note}
  {(unpublished)}\BibitemShut {NoStop}%
\bibitem [{\citenamefont {{Nayak}}\ and\ \citenamefont
  {{Wilczek}}(1995)}]{NayakWilczekDDW}%
  \BibitemOpen
  \bibfield  {author} {\bibinfo {author} {\bibfnamefont {C.}~\bibnamefont
  {{Nayak}}}\ and\ \bibinfo {author} {\bibfnamefont {F.}~\bibnamefont
  {{Wilczek}}},\ }\href@noop {} {\  (\bibinfo {year} {1995})},\ \Eprint
  {http://arxiv.org/abs/arXiv:cond-mat/9510132} {arXiv:cond-mat/9510132}
  \BibitemShut {NoStop}%
\bibitem [{\citenamefont {Varma}(1997)}]{VarmaNFLPairing}%
  \BibitemOpen
  \bibfield  {author} {\bibinfo {author} {\bibfnamefont {C.~M.}\ \bibnamefont
  {Varma}},\ }\href {\doibase 10.1103/PhysRevB.55.14554} {\bibfield  {journal}
  {\bibinfo  {journal} {Phys. Rev. B}\ }\textbf {\bibinfo {volume} {55}},\
  \bibinfo {pages} {14554} (\bibinfo {year} {1997})}\BibitemShut {NoStop}%
\bibitem [{\citenamefont {Varma}(1999)}]{VarmaPseudoGap}%
  \BibitemOpen
  \bibfield  {author} {\bibinfo {author} {\bibfnamefont {C.~M.}\ \bibnamefont
  {Varma}},\ }\href {\doibase 10.1103/PhysRevLett.83.3538} {\bibfield
  {journal} {\bibinfo  {journal} {Phys. Rev. Lett.}\ }\textbf {\bibinfo
  {volume} {83}},\ \bibinfo {pages} {3538} (\bibinfo {year}
  {1999})}\BibitemShut {NoStop}%
\bibitem [{\citenamefont {Nayak}(2000)}]{NayakDensityWave}%
  \BibitemOpen
  \bibfield  {author} {\bibinfo {author} {\bibfnamefont {C.}~\bibnamefont
  {Nayak}},\ }\href {\doibase 10.1103/PhysRevB.62.4880} {\bibfield  {journal}
  {\bibinfo  {journal} {Phys. Rev. B}\ }\textbf {\bibinfo {volume} {62}},\
  \bibinfo {pages} {4880} (\bibinfo {year} {2000})}\BibitemShut {NoStop}%
\bibitem [{\citenamefont {Chakravarty}\ \emph {et~al.}(2001)\citenamefont
  {Chakravarty}, \citenamefont {Laughlin}, \citenamefont {Morr},\ and\
  \citenamefont {Nayak}}]{ChakravartyCuprates}%
  \BibitemOpen
  \bibfield  {author} {\bibinfo {author} {\bibfnamefont {S.}~\bibnamefont
  {Chakravarty}}, \bibinfo {author} {\bibfnamefont {R.~B.}\ \bibnamefont
  {Laughlin}}, \bibinfo {author} {\bibfnamefont {D.~K.}\ \bibnamefont {Morr}},
  \ and\ \bibinfo {author} {\bibfnamefont {C.}~\bibnamefont {Nayak}},\ }\href
  {\doibase 10.1103/PhysRevB.63.094503} {\bibfield  {journal} {\bibinfo
  {journal} {Phys. Rev. B}\ }\textbf {\bibinfo {volume} {63}},\ \bibinfo
  {pages} {094503} (\bibinfo {year} {2001})}\BibitemShut {NoStop}%
\end{thebibliography}%


\newpage
\begin{widetext}
\begin{appendix}
\renewcommand{\thesection}{$\text{S}$}
\setcounter{equation}{0}
\renewcommand{\thefigure}{$\text{S}$\arabic{figure}}
\setcounter{figure}{0}

\section{SUPPLEMENTARY MATERIAL}
\section{VARIATIONAL PHASE DIAGRAM}
We consider the extended Bose-Hubbard model on the inverted triangular lattice
\begin{equation}
H=t\sum_{\langle i,j\rangle}\left(b_i^{\dag}b_{j}+{\rm h.c.}\right)+\frac{U}{2}\sum_i(n_i-n_0)^2+V\sum_{\langle i,j\rangle}(n_i-n_0)(n_j-n_0)=H_0+H_U+H_V.
\end{equation}
Using mean-field wave-functions for the chiral superfluid and the chiral Mott insulator, we obtain the schematic phase diagram shown in Fig.~1(c) of the main text.
In the following we discuss the mean-field wave functions and their corresponding energies.

\subsection{Variational analysis  of the chiral Mott Insulator}
The variational wave-function for the chiral Mott insulator
is
\begin{equation}
\ket{\chi_{MI}}=\mathcal{N}_{\Delta}^{-1/2}e^{\Delta\sum_{\bs k}\varphi_{\bs k}b_{\bs k}^{\dag}b_{\bs k}}|MI\rangle=\mathcal{N}_{\Delta}^{-1/2}\left(1+\Delta A+\frac{\Delta^2}{2}A^2+\dots\right)|MI\rangle
\end{equation}
where  $A = \sum_{\bs k}\varphi_{\bs k}b_{\bs k}^{\dag}b_{\bs k}$, and  the normalization is
\begin{equation}
\mathcal{N}_\Delta=\langle MI|e^{2\Delta\sum_{\bs k}\varphi_{\bs k}b_{\bs k}^{\dag}b_{\bs k}}|MI\rangle.
\end{equation}
The function $\varphi_{\bs k}$ incorporates the chiral symmetry breaking and should be odd under parity,
$\varphi_{\bs k}=-\varphi_{-{\bs k}}$;
 the simplest form (which we will eventually use) is obtained if one considers imaginary hopping around the triangles,
$\varphi_{\bs k}^{(0)}=2\sum_{i}\sin({\bs k}\cdot{\bs a}_i)$, 
 but for the present we keep $\varphi_{\bs k}$ unspecified.
In our convention, the energy of the trivial MI vanishes, 
$\langle MI|H_0|MI\rangle=\langle MI|H_{\rm int}|MI\rangle=0$ and so $\mathcal{N}_\Delta$ does not contribute to the variational energy upto quadratic order in $\Delta^2$. 
For the kinetic energy we evaluate
\begin{equation}
E_{\rm kin}=\Delta^2\left(\langle A H_0 A\rangle+\frac{1}{2}\langle H_0 A^2\rangle+\frac{1}{2}\langle A^2 H_0 \rangle\right)=2\Delta^2\langle H_0 A^2\rangle.
\end{equation}
Here, the expectation value is with respect to the trivial Mott insulator and we have used the fact that $A$ commutes with $H_0$.
Writing $A$ in real space, we find
\begin{equation}
A=\sum_{\bs k}\varphi_{\bs k}b^{\dag}_{\bs k}b_{\bs k}=\sum_{i,j}b_i^{\dag}b_j\left(\frac{1}{N}\sum_{\bs k}\varphi_{\bs k}e^{-i{\bs k}\cdot({\bs r}_i-{\bs r}_j)}\right)=\sum_{i,j}w_{ij}b_i^{\dag}b_j.
\end{equation}
Because of the odd-parity constraint on $\varphi_{\bs k}$, the hopping amplitude is purely imaginary, $w_{ij}^*=-w_{ij}$ and $w_{ii}=0$. In order to evaluate the kinetic energy, we need the following expectation value
\begin{eqnarray}
\langle b_i^{\dag}b_{i+\delta}b_j^{\dag}b_kb_l^{\dag}b_m\rangle&=&\delta_{m,i}\delta_{j,i+\delta}\delta_{kl}\langle b_i^{\dag}b_{i+\delta}b_{i+\delta}^{\dag}b_kb_k^{\dag}b_i\rangle+\delta_{i,k}\delta_{l,i+\delta}\delta_{m,j}\langle b_i^{\dag}b_{i+\delta}b_j^{\dag}b_ib_{i+\delta}^{\dag}b_j\rangle\\
&=&\delta_{m,i}\delta_{j,i+\delta}\delta_{kl}n_0(n_0+1)^2+\delta_{i,k}\delta_{l,i+\delta}\delta_{m,j}n_0^2(1+n_0)
\label{eq:exp4b}
\end{eqnarray}
where as before $\langle\dots\rangle$ denote expectation values with respect to the trivial Mott state and we used the fact that $j\neq k$ and $l\neq m$. Because of this constraint, the bosons need to hop around a triangle (but not necessary an elementary triangle, although this will be the case for the simple nearest-neighbor choice  $\varphi_{\bk}^{(0)}$). Using Eq.~\eqref{eq:exp4b}, we find for the kinetic energy
\begin{equation}
E_{\rm kin}=2\Delta^2 t \sum_{i,\delta}\sum_{k\neq i,i+\delta}w_{i+\delta,k}w_{k,i}n_0(n_0+1)(2n_0+1)=2\Delta^2 zt\left(\frac{1}{N}\sum_{\bs k}\varphi_{\bs k}^2\gamma_{\bs k}\right)n_0(n_0+1)(2n_0+1)N.
\end{equation}
Here, we have introduced the single-particle dispersion from nearest-neighbor hopping on the triangular lattice:
\begin{equation}
\varepsilon_{\bs k}/t=z\gamma_{\bs k}=\sum_{\bs \delta}e^{i{\bs k}\cdot{\bs \delta}}=2\sum_i\cos({\bs k}\cdot{\bs a}_1).
\end{equation}
where $z=6$ is the coordination number. For the simplest {\it ansatz}, $\varphi_{\bs k}=\varphi_{\bs k}^{(0)}$, we have
$\frac{1}{N}\sum_{\bs k}(\varphi_{\bs k}^{(0)})^2\gamma_{\bs k}=-2
$ so that the variational kinetic energy is
\begin{equation}
E_{\rm kin}^{(0)}=-4zt\Delta^2n_0(n_0+1)(2n_0+1)N.
\end{equation}
To order $\Delta^2$ , the onsite repulsion contributes
$E_{U}=\Delta^2\langle A H_{U} A\rangle$ to the variational energy. 
To evaluate $E_U$, we need to compute
\begin{equation}
\sum_k\langle b_i^{\dag}b_j (n_k-n_0)^2 b_l^{\dag}b_m\rangle=\delta_{jl}\delta_{im}\langle b_i^{\dag}b_j[(n_i-n_0)^2+(n_j-n_0)^2]b^{\dag}_jb_i\rangle=2\delta_{jl}\delta_{im}n_0(n_0+1).
\end{equation}
It follows that
\begin{equation}
E_{U}=2\frac{U}{2}\Delta^2n_0(n_0+1)\sum_{i,j}\frac{1}{N^2}\sum_{\bs k}\sum_{\bs q}\varphi_{\bs k}\varphi_{\bs q}e^{-i{\bs r}_i\cdot({\bs k}-{\bs q})}e^{i{\bs r}_j\cdot({\bs k}-{\bs q})}=U\Delta^2n_0(n_0+1)N\left(\frac{1}{N}\sum_{\bs k}\varphi_{\bs k}^2\right)
\end{equation}
Using
$\frac{1}{N}\sum_{\bs k}(\varphi^{(0)}_{\bs k})^2=z=6
$, 
we find for the potential energy
\begin{equation}
E_{U}^{(0)}=NzU\Delta^2n_0(n_0+1).
\end{equation}
To order $\Delta^2$ , the nearest-neighbor repulsion contributes
$E_{V}=\Delta^2\langle A H_{V} A\rangle.$ 
Hence, we need to evaluate ($l\neq k$, $m\neq n$) 
\begin{equation}
\langle b_l^{\dag}b_k (n_i-n_0)(n_j-n_0) b_m^{\dag}b_n\rangle=-\delta_{ln}\delta_{km}\left(\delta_{nj}\delta_{mi}+\delta_{mj}\delta_{ni}\right)n_0(n_0+1).
\end{equation}
It follows that
\begin{eqnarray}
\langle H_V\rangle&=&-\Delta^2 \frac{V}{2}n_0(n_0+1)\sum_{i}\sum_{\bs \delta}\frac{1}{N^2}\sum_{\bs k}\sum_{\bs q}\varphi_{\bs k}\varphi_{\bs q}\left[e^{-i{\bs \delta}\cdot({\bs k}-{\bs q})}+e^{i{\bs \delta}\cdot({\bs k}-{\bs q})}\right]\\
&=&-N\Delta^2\frac{V}{t}n_0(n_0+1)\frac{1}{N^2}\sum_{{\bs k},{\bs q}}\varphi_{\bs k}\varphi_{\bs q}\varepsilon_{{\bs k}-{\bs q}}.
\end{eqnarray}
Using $\frac{1}{N^2}\sum_{{\bs k},{\bs q}}\varphi_{\bs k}^{(0)}\varphi_{\bs q}^{(0)}\varepsilon_{{\bs k}-{\bs q}}=zt
$ so that
\begin{equation}
E_V^{(0)}=-N\Delta^2Vzn_0(n_0+1).
\end{equation}
All together, we find for the variational energy in the chiral Mott phase
\begin{equation}
E_{\chi_{MI}}^{(0)}=N\Delta^2z n_0(n_0+1)\left[-4t(2n_0+1)+(U-V)\right]
\end{equation}
from which the critical hopping  for the $MI-\chi_{MI}$ transition is found
\begin{equation}
t_{\chi_{MI}}^{(0)}=\frac{U-V}{4(2n_0+1)}.
\end{equation}
Note that the effective interaction which enters this expression is $\tilde{U}=U-V$.
%
%
\subsubsection{Momentum distribution in  the Chiral MI}
The momentum distribution is calculated as
\begin{equation}
\langle n_{\bs k}\rangle_{\chi_{MI}}=\langle n_{\bs k}\rangle+2\Delta\langle n_{\bs k}A\rangle+ 2\Delta^2\left(\langle n_{\bs k}A^2\rangle-\langle n_{\bs k}\rangle\langle A^2\rangle\right)+\cdots
\end{equation}
where as before the expectation values on the r.h.s are taken with respect to the trivial MI. Unlike for the variational energy, the normalization contributes in order $\Delta^2$. However, we can focus on the smallest order in $\Delta$ which is linear.
We have
\begin{equation}
\langle n_{\bs k}\rangle=\frac{1}{N}\sum_{i,j}e^{i{\bs k}\cdot({\bs r}_i-{\bs r}_j)}\langle b_i^{\dag}b_j\rangle=n_0.
\end{equation}
Furthermore, using similar manipulations as above, we find
\begin{equation}
\langle n_{\bs k}A\rangle=n_0(1+n_0)\varphi_{\bs k}.
\end{equation}
So the momentum distribution in linear order in $\Delta$ is modified as
\begin{equation}
\langle n_{\bs k}\rangle=n_0[1+2\Delta (1+n_0) \varphi_{\bs k}].
\end{equation}
\subsection{Variational analysis of the chiral superfluid}
The variational wave function for the chiral superfluid is
\begin{eqnarray}
|\chi_{SF}\rangle&=&\mathcal{N}_\psi^{-1/2}\exp\left({\psi \frac{b_{\bs K}^{\dag}}{\sqrt{n_0+1}}+\psi^*\frac{b_{\bs K}}{\sqrt{n_0}}}\right)|MI\rangle\\
&=&\mathcal{N}_\psi^{-1/2}\left\{1+\psi \frac{b_{\bs K}^{\dag}}{\sqrt{n_0+1}}+\psi^*\frac{b_{\bs K}}{\sqrt{n_0}}+\frac{1}{2}\left[\frac{\psi^2}{n_0+1}(b_{\bs K}^{\dag})^2+\frac{(\psi^*)^2}{n_0}b_{\bs K}^2+\frac{2|\psi|^2}{\sqrt{n_0(1+n_0)}}(b_{\bs K}^{\dag}b_{\bs K}+\frac{1}{2})\right]+\dots\right\}|MI\rangle.\nonumber
\end{eqnarray}
Once again, the normalization $\mathcal{N}_\psi$ only contributes at quartic and higher order to the variational energy and can therefore be ignored. 
To order $|\psi|^2$, the kinetic energy is
\begin{equation}
E_{\rm kin}=|\psi|^2\left(\frac{1}{n_0+1}\langle b_{\bs K}H_{\rm kin}b_{\bs K}^{\dag}\rangle+\frac{1}{n_0}\langle b_{\bs K}^{\dag} H_{\rm kin}b_{\bs K}\rangle+\frac{2}{\sqrt{n_0(1+n_0)}}\langle H_{\rm kin}(n_{\bs K}+1/2)\rangle\right).
\end{equation}
The first term is 
\begin{equation}
\langle b_{\bs K}H_{\rm kin}b_{\bs K}^{\dag}\rangle=-t\sum_{lm}\sum_{i\delta}\langle b_lb_i^{\dag}b_{i+\delta}b_m^{\dag}\rangle e^{i{\bs K}\cdot({\bs r}_l-{\bs r}_m)}=(n_0+1)^2N\varepsilon({\bs K}).
\end{equation}
Similarly,
\begin{eqnarray}
\langle b_{\bs K}^{\dag} H_{\rm kin}b_{\bs K}\rangle&=&n_0^2N\varepsilon({\bs K})\\
\langle H_{\rm kin}(n_{\bs K}+1/2)\rangle&=&n_0(1+n_0)N\varepsilon({\bs K}).
\end{eqnarray}
All together
\begin{equation}
E_{\rm kin}=|\psi|^2N\varepsilon({\bs K})(\sqrt{n_0+1}+\sqrt{n_0})^2.
\end{equation}
To order $|\psi|^2$, the onsite repulsion contributes
\begin{equation}
E_{U}=|\psi|^2\left[\frac{1}{n_0+1}\langle b_{\bs K}H_{U}b_{\bs K}^{\dag}\rangle+\frac{1}{n_0}\langle b_{\bs K}^{\dag} H_{U}b_{\bs K}\rangle+\frac{1}{\sqrt{n_0(1+n_0)}}\left(\langle H_{U}(n_{\bs K}+1/2)\rangle+\langle (n_{\bs K}+1/2)H_{U}\rangle\right)\right].
\end{equation}
The first term is
\begin{equation}
\langle b_{\bs K}H_{U}b_{\bs K}^{\dag}\rangle=\frac{U}{2}\sum_{l,m}\langle b_l(n_i-n_0)(n_i-n_0)b_m^{\dag}\rangle e^{i{\bs K}\cdot({\bs r}_l-{\bs r}_m)}=\frac{U}{2}(n_0+1)N.
\end{equation}
Similarly, 
\begin{equation}
\langle b_{\bs K}^{\dag} H_{U}b_{\bs K}\rangle=\frac{U}{2}n_0N.
\end{equation}
The third term vanishes, so all together we have
\begin{equation}
E_U=|\psi|^2NU.
\end{equation}
The nearest-neighbor interaction does not contribute to the energy in order $|\psi|^2$, i.e. $E_V=0.$ 
This is seen for example by considering
\begin{equation}
\langle b_{\bs K}H_{V}b_{\bs K}^{\dag}\rangle=\frac{V}{2}\sum_{lm}\sum_{i,\delta}\langle b_l (n_i-n_0)(n_{i+\delta}-n_0)b_m^{\dag}\rangle=0.
\end{equation}

The transition from the trivial Mott to the chiral superfluid can be obtained from the well-known (mean-field) expression [S.~D.~Huber, E. Altman, H. P. Buchler, and G. Blatter, Phys. Rev. B {\bf 75}, 085106 (2007)]: 
\begin{equation}
U_{MI}=-\varepsilon({\bs K})(\sqrt{n_0+1}+\sqrt{n_0})^2=\frac{zt}{2}(\sqrt{n_0+1}+\sqrt{n_0})^2.
\end{equation}
From this, we obtain the nearest-neighbor interaction for which the chiral MI and chiral SF are degenerate,
\begin{equation}
V_c(U,n_0)=U\frac{3(\sqrt{n_0+1}+\sqrt{n_0})^2-4(2n_0+1)}{3(\sqrt{n_0+1}+\sqrt{n_0})^2}.
\end{equation}
In the large $n_0$ limit, we have
\begin{equation}
V_c(U,n_0\rightarrow\infty)=\frac{U}{3}.
\end{equation}

\section{EFFECTIVE HOPPING AND NEAREST-NEIGHBOR REPULSION}
An inverted hopping matrix element can be produced quite naturally for atoms which reside in upper bands. Let us consider a system with at least two sites denoted by $\a=1,2$ per unit cell. We take the energy of the atom localized on sites $\a=1$ to be higher by $\Delta$ than atoms on sites $\a=2$. Moreover a site of type $2$ is situated between every two sites of type $1$. This is the case for example if the lower energy sites form a Kagom\'e lattice whereas the higher energy sites reside at the centers of the hexagons and form a triangular lattice. Hopping between triangular lattice sites takes place through an intermediate Kagom\'e site of lower energy. 

 If the tunneling between the sites $t_{12}$ is much smaller than $\D$ then the Bloch wave function in the upper band is weighted mostly on sites of type $\a=1$. In this case the effective tunneling between the higher energy sites can be calculated in perturbation theory as $t= -t_{12}^2/\Delta$. The sign is opposite of regular hopping, i.e. favors a $\pi$ phase change, because of the hopping through a {\em lower} energy intermediate state. 
 
Nearest neighbor interactions in the upper band are produced as follows. The excess energy for having two neighboring upper band sites ($\a=1$) occupied stems from the interaction when they both virtually occupy the intermediate lower energy site. Within fourth order perturbation theory this interaction is given by
\be
V=8{t_{12}^4\over \D^2 (\D-U_2)}= 8 t^2/(\D-U_2) ,
\ee 
where $U_2$ is the on-site interaction on the intermediate site. Note that $V$ is repulsive if $U_2<\Delta$. This interaction can be significant in spite of being fourth order because $\D-U_2$ appearing in the energy denominator can in principle be made small (i.e. near resonance). We note however that if it is too small then the effective interaction should be calculated non perturbatively. The condition for the above perturbation theory to hold is $t\ll\sqrt{t_{12}(\D-U_2)}$.

\section{FINITE ENTANGLEMENT SCALING FOR 4-LEG LADDER}
\begin{figure}[t]
\includegraphics[width=0.45\columnwidth]{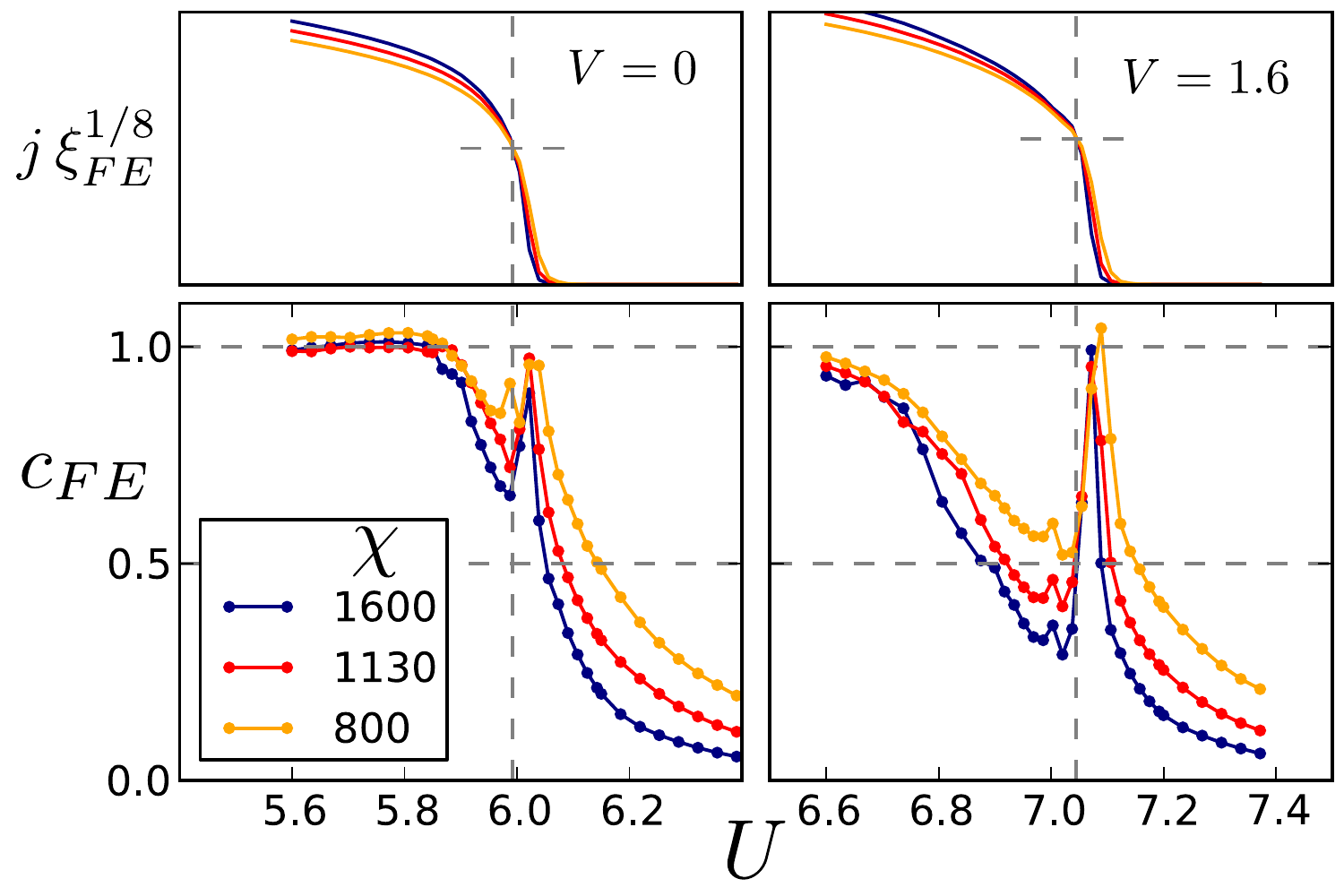}
\caption{\label{4legfes} Finite entanglement scaling of the bond current $j$ (upper panel) and central charge $c$ (lower panel) for a 4-leg ladder. Left and right panels are for data at $V=0$ and $V = 1.6$ respectively.}
\end{figure}
Here we include some additional details on the iDMRG analysis of the 4-leg ladder. 
Data was collected at several bond dimensions up through $\chi = 1600$.
Analogous to finite size scaling, we can obtain finite entanglement estimates of the critical exponents and  central charge $c$ through the dependence of observables on the bond dimension $\chi$.
Note that the critical exponents will always be those of a 1-D system, as they implicitly are due to the fluctuations at long distances along the cylinder, at a scale much larger than the circumference. 
Nevertheless, for our purposes we can determine if the state  a) breaks the chiral symmetry and b) is gapped.

	To investigate the chiral symmetry breaking, we measure the scaling of the $U(1)$ bond current $j$ with finite entanglement correlation length $\xi_{FE}$.
Assuming Ising criticality, the combination $j \xi_{FE}^{1/8}$ is dimensionless, so we should find a crossing at the critical point $U_c$. 
As shown in the top panels of Fig. \ref{4legfes}, for $V = 0, 1.6$ respectively, the crossing allows us to measure $U_c$, shown by the vertical line. We fine $U_c = 5.99, 7.06$ respectively.

	The existence of a gap can be determined by measuring the  central charge $c$.
It is known the the entanglement entropy scales as $S = \frac{c}{6} \log(\xi_{FE})$,\cite{pollmann2009} from which we can estimate $c$.
Consider first $V=1.6$, for which the results are quite clear. 
At $U_c$, the FES estimate of $c$ is  $c_{FE} = \frac{1}{2}$, independent of $\chi$; this shows we are in the scaling regime of an Ising critical point.
For $U_c \ll U $, the system is in a Mott insulating phase; $c_{FE}$ is both less than $c = 1/2$ and renormalizing downward with the bond dimension $\chi$, showing that the region is gapped. 
The peak in $c_{FE}$ for $U_c \lesssim U$ is an FES artifact, similar to the finite-size shift in $T_c$ familiar from finite-size scaling.
In the region $U < U_c$, the putative $\chi$MI, we see that $c_{FE}$ strongly renormalizes downward, showing the state is gapped.
For some $U_{KT} < 6.6$ (not shown), we find $c_{FE} = 1$, showing the BKT transition into the $c = 1$ XY phase.
A more involved analysis of the superfluid stiffness is required to determine the precise location of $U_{KT}$.
In summary, there is definitive evidence for a $\chi$MI phase at $V = 1.6$.

	The evidence of the $\chi$MI is less clear at $V = 0$. The system is certainly in the XY phase for $U < 5.7$, and has Ising critical exponents at $U_c$, as evidenced by the scaling of $j \xi_{FE}^{1/8}$.
There is a small region $5.8 < U < U_c$ in which $c_{FE}$ is renormalizing downward, but since the estimate of $c_{FE}$ has not converged at $U_c$, we can conclude the data is not fully in the scaling regime at $\chi= 1600$.
Nevertheless, the apparent dip in $c_{FE}$ strongly suggests the BKT and Ising transitions are split.

	Regardless of the conclusion for the 4-leg ladder, we can conclude little about the existence of the $\chi MI$ phase at $V=0$ in the isotropic case; this would require a detailed scaling analysis as the circumference of the cylinder is increased.
	\end{appendix}
	\end{widetext}
\end{document}